\documentclass[twocolumn,showpacs]{revtex4}
\usepackage[xdvi]{graphicx}%
\usepackage{dcolumn}
\usepackage{amsmath}
\makeatletter

\def\btt#1{\texttt{\@backslashchar#1}}%
\DeclareRobustCommand\bblash{\btt{\@backslashchar}}%
\makeatother

\newcommand{\bra}{\left\langle}
\newcommand{\ket}{\right\rangle}
\newcommand{\kets}{\right\rangle_{\rm s}}

\newcommand{\e}{{\rm e}}

\newcommand{\jp}{j_{\rm p}}
\newcommand{\jt}{j_{\rm t}}

\newcommand{\veceta}{{\boldsymbol \eta}}

\begin{document}

\title{Fluctuation-dissipation relations outside the linear response 
regime  in a two-dimensional driven lattice gas along the direction  
transverse to the driving force} 
\author{Kumiko Hayashi}
\affiliation
{Department of Pure and Applied Sciences,
University of Tokyo, Komaba, Tokyo 153-8902, Japan}

\date{\today}

\begin{abstract}
We performed numerical experiments on a two-dimensional driven 
lattice gas, which constitutes a simple stochastic nonequilibrium 
many-body model. In this model, focusing on the behavior along the 
direction transverse to the external driving force, we numerically 
measure transport coefficients and dynamical fluctuations outside 
the linear response regime far from equilibrium. Using these 
quantities, we find the validity of 
the Einstein relation, the Green-Kubo relation and the 
fluctuation-response relation.  
\end{abstract}

\maketitle

 
In order to construct a theoretical framework of nonequilibrium 
statistical mechanics,  we  seek  new universal relations characterizing 
nonequilibrium steady states (NESS) far from equilibrium. For this 
purpose, we carry out numerical computations through which we study 
the validity of certain fluctuation-dissipation relations. In this 
paper, we consider one such model,  a simple stochastic nonequilibrium 
many-body model which is called a 'driven lattice gas (DLG)' 
\cite{dlg1,dlg3}.  

\paragraph*{Model:} 

Let $\eta_i$ be an occupation variable defined on each site 
$i=(i_x,i_y)$ of a two-dimensional square lattice $\{(i_x,i_y)| 0 
\le i_x \le L,  0\le i_y \le L\}$.  The  variable $\eta_i$ is 1 if 
the $i$th site is occupied by a particle, and 0 if it is unoccupied. 
Periodic boundary conditions are imposed by setting $\eta_i=\eta_j$, 
where $j=(L,i_y)$ in the case $i_x=0$, and  $\eta_i=\eta_j$, where 
$j=(i_x,L)$ in the case $i_y=0$. The array of all occupation 
variables, $\{\eta_i\}$, is denoted  $\veceta$ and called the 
``configuration''.

The time evolution of $\veceta$ is described by the following
rule: At each time step, randomly  choose  a nearest-neighbor pair 
$\bra i,j\ket$, and exchange the values of $\eta_i$ and $\eta_j$ 
with the  probability  $c(i,j;\veceta)=\{1+\exp[\beta Q(\veceta\to 
\veceta^{ij})]\}^{-1}$,  where $\veceta^{ij}$ is the configuration 
obtained from $\veceta$ through this exchange, and $\beta=1/T$ is the 
inverse temperature with the Boltzmann constant  set to unity.  
$Q(\veceta\to \veceta^{ij})$ represents the heat absorbed from the 
heat bath as a result of the configuration change $\veceta\to
\veceta^{ij}$. The total particle number, $N=\sum_{i}\eta_i$, is 
conserved throughout the time evolution. The density $\rho=N/L^2$ is 
a parameter of the model. Hereafter,  we regard the unit of time to be 
$L^2$ time steps, which is  the number of time steps for which an 
arbitrary site is chosen once  on average. We refer to this time 
as 1 MCS  (Monte Carlo step per site).

In this paper, we study a two-dimensional DLG with
\begin{equation}
Q(\veceta\to\veceta')
\equiv H_0(\veceta')-H_0(\veceta) - E \jp(\veceta\to\veceta'),
\label{heat}
\end{equation}
where $E$ is an external driving force, and $H_0(\veceta)$ describes 
an interaction between particles, written  $H_0(\veceta)\equiv 
-\sum_{ \bra i,j \ket} \eta_i \eta_j$, where $\bra i,j \ket$ denotes 
a nearest-neighbor pair. The quantity $\jp(\veceta\to\veceta')$ is 
the spatially-averaged current, that is, the  net number of particles 
flowing  in the $x$ direction: $\jp(\veceta\to\veceta')
\equiv \sum_i[\eta_i(1-\eta'_{i})\eta'_{i+(1,0)}(1-\eta_{i+(1,0)})
-\eta'_i(1-\eta_i)\eta_{i+(1,0)}(1-\eta'_{i+(1,0)})]$.     
In this study, we fix $\beta=0.5$ in order for the system to be far 
from the critical region (Note that the critical temperature of the 
model with $E=0$ is $\beta_{\rm c}=1.76$.), 
and choose large values of $E$ in order 
for the system to be far from equilibrium.  

\paragraph*{Our aim:} 


In DLGs, regarded as one of the simplest classes of 
nonequilibrium models,  statistical properties of NESS have been  
investigated from various points of view.  Among them, there is an 
interesting report that a large deviation functional of density 
fluctuations is shape dependent \cite{eyink}. 
In a two-dimensional DLG,  although   
the general properties of fluctuations 
are quite different from those of equilibrium states,  it was found 
numerically that a fluctuation relation 
holds \cite{HSI}, where we consider only properties along the direction 
transverse to the external force $E$. This  fluctuation relation is 
a relation among  density fluctuations, 
the chemical potential \cite{chemi,sst}, and the temperature of the  
environment.


In equilibrium cases, the fluctuation relation is closely related to 
fluctuation-dissipation relations, which relate  dynamical 
properties of equilibrium fluctuations with transport  
properties in the linear response regime \cite{HSVI}.   Then noting 
that the  fluctuation relation holds in the DLG even far from 
equilibrium \cite{HSI}, we wish to also investigate the validity of 
fluctuation-dissipation relations far from equilibrium, and determine 
if their equilibrium forms hold here as well.

In order to obtain such relations, we directly measure transport 
coefficients and dynamical fluctuations in the direction transverse 
to the driving force $E$ in the two-dimensional DLG investigated above. 
In spite of the fact that  these measured values differ from those 
for the equilibrium state($E=0$),   we numerically find that three 
fluctuation-dissipation relations, the Einstein relation, 
the fluctuation-response relation, and the Green-Kubo relation, seem 
to be valid even for NESS far from equilibrium.

\paragraph*{Einstein relation:}  

%
%
\begin{figure}
\begin{center}
\includegraphics[width=8cm]{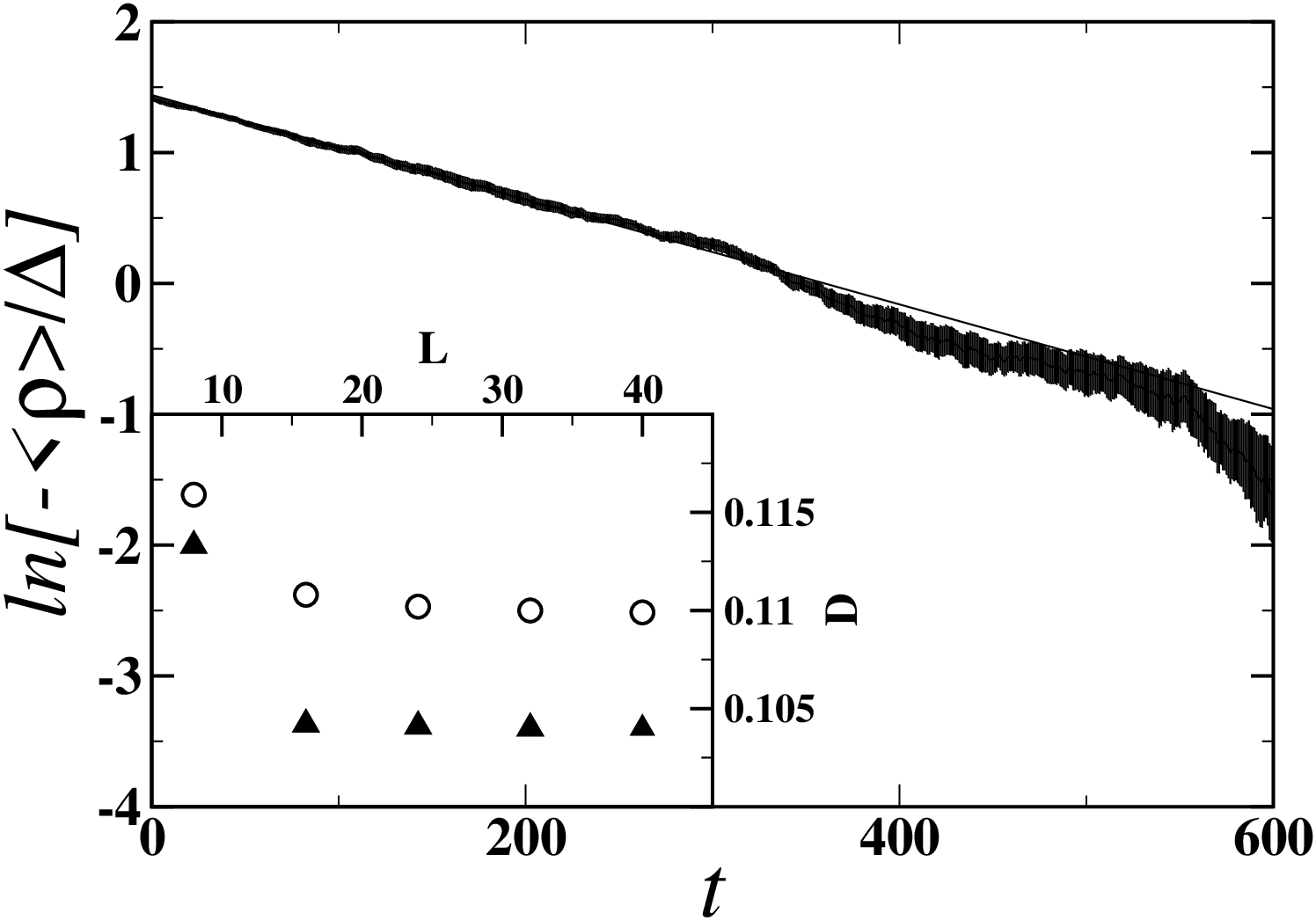}
\end{center}
\caption{The quantity $\ln[-\bra\hat\rho(t)\ket_{E}^{V}/\Delta]$ 
plotted as a function of $t$ in the case $(\rho,E)=(0.5,10)$,  
with $\Delta=-0.2$, $L=32$. From the slope of the line,  
$-0.004t+1.44$, $D$ is estimated as $0.004=D(2\pi/L)^2$. The 
inset displays the $L$ dependence of $D$. The triangles and circles  
correspond to $(\rho,E)=(0.5,10)$ and  $(0.5,0)$, respectively.
}
\label{fig:exp}
\end{figure}

In the linear response regime near $E=0$, the Einstein relation 
for interacting many-body systems is written  
\begin{equation}
D\chi=\sigma T, 
\label{ein}
\end{equation}
where $D$ is the  density diffusion constant, $\chi$ is the intensity 
of density fluctuations, and $\sigma$ is  the conductivity \cite{kubo}.  

In order to  investigate the validity of (\ref{ein}) in the direction 
transverse  to the external driving force $E$ (the $y$-direction),  
we first need to define the 
density diffusion constant $D$ as the  coefficient of the  
diffusion term in the evolution equation describing the   averaged 
behavior of a density field in this direction.  In  this paper, 
we consider the 
case in which we have prepared as 
the initial state, a steady state 
under the  perturbation potential 
\begin{equation}
V(i_y)=\Delta\sin\frac{2\pi i_y}{L}  
\label{pote}
\end{equation}
which is obtained  by adding $\sum_i \eta_i V(i_y)$  
to $H_0(\veceta)$. Then, we remove 
$V(i_y)$  at $t=0$ in order to measure the relaxation of the density 
field $\hat\rho(t)$. The function $\hat\rho(t)$ is the Fourier transform 
of the coarse-grained density  $\rho(i_y)$. These quantities 
are defined as  $\hat\rho(t)\equiv\sum_{i_y=1}^{L}\rho(i_y)
\sin\frac{2\pi i_y}{L}$, and $\rho(i_y)\equiv\frac{1}{L}  
\sum_{i\in\Omega_{i_y}} \eta_i$, where $\Omega_{i_y}=\{(i_x,i_y) |  
1\le i_x\le L, i_y\}$.  

In Fig. \ref{fig:exp},  choosing $\Delta$ as a sufficiently small 
value, $\ln(-\bra\hat\rho(t)\ket_{E}^{V}/\Delta)$ is 
plotted as a function of $t$ in the case $(\rho,E)=(0.5,10)$ with  
$L=32$  and $\Delta=-0.2$.  Here, $\bra\ \ket_{E}^{V}$ 
represents the statistical average under the relaxation process. 
Because exponentially decaying behavior of $\bra \hat \rho(t) 
\ket_{E}^{V}$ is observed, $D$ can be estimated from  
the form 
\begin{equation} 
\bra \hat \rho(t) \ket_{E}^{V}
= {\rm const.}\e^{- D \left(\frac{2\pi}{L}\right)^2 t}. 
\label{diff}
\end{equation}
In the inset of Fig. \ref{fig:exp},  $D$ is  plotted as a function 
of the system  size $L$ in the cases $E=0$ and $E=10$, with  
$\rho=0.5$. Because both  values of $D$ seem to  
converge, we conclude that the size $L=40$ can be regarded as  
sufficiently large  to study  the statistical properties 
of macroscopic quantities in our  model. It is important to note here 
that  the values of $D$ in the case $E=10$ are different from those 
in the case $E=0$.

\begin{figure}
\begin{center}
\includegraphics[width=6cm]{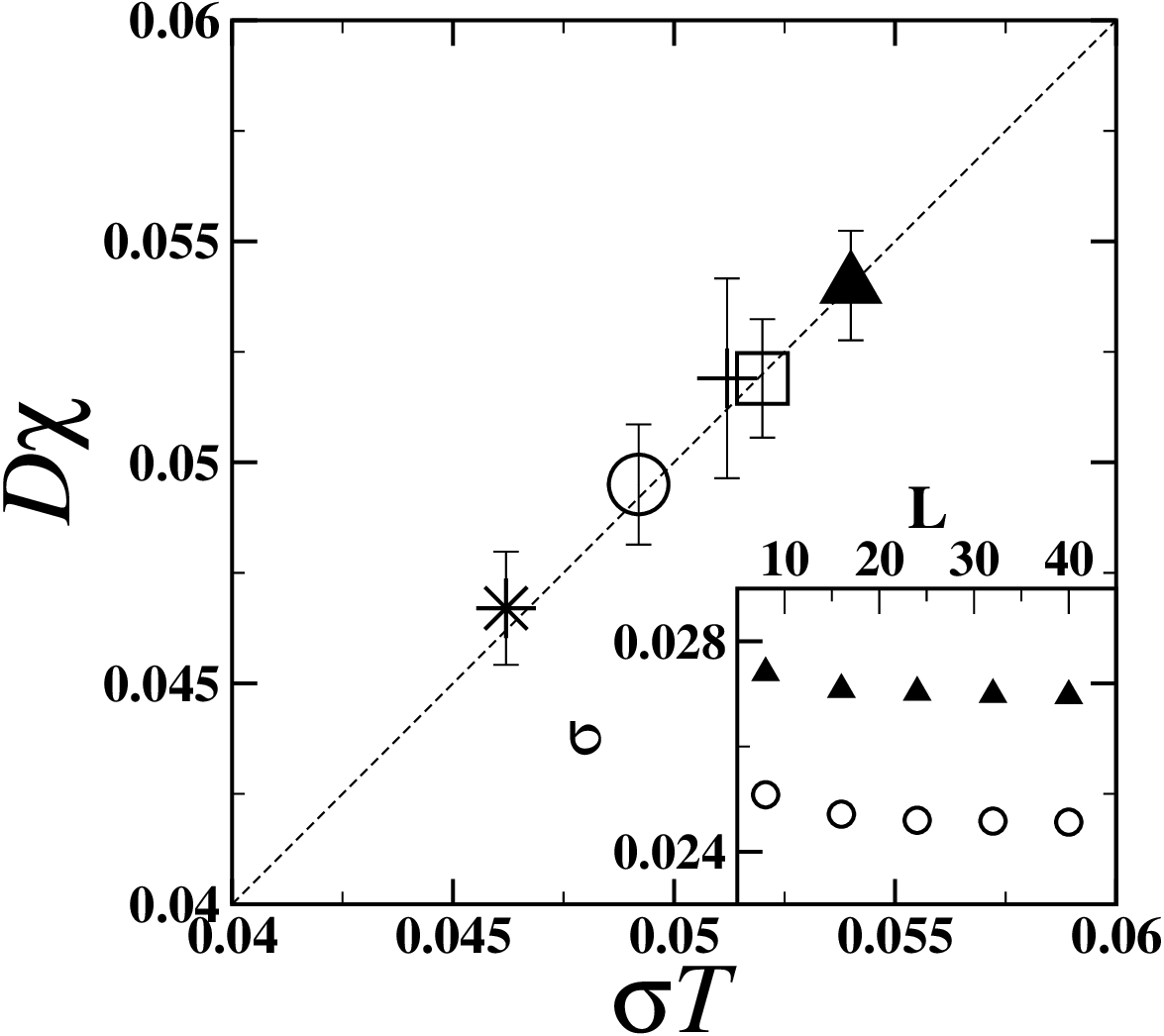}
\end{center}
\caption{$\sigma T$ as a function of $D\chi$. 
The triangle, square, star, plus, and circle correspond to 
$(\rho,E)=(0.5,10),(0.4,10),(0.3,10),(0.5,3)$ and $(0.5,0)$. 
The error bars  represent the statistical errors arising from the 
fitting when $D$ is measured. The thin-dotted line represents 
$D\chi=\sigma T$.  The inset displays the $L$ dependence of 
$\sigma$. The triangles and circles correspond to $(\rho,E)=(0.5,10)$ 
and  $(0.5,0)$, respectively. 
}
\label{fig:ein}
\end{figure}

Next, we define the  conductivity $\sigma$ by adding a sufficiently 
small perturbativeg driving force $\epsilon$ in the $y$-direction. 
This is realized by adding the term $\epsilon \jt(\veceta \to \veceta')$ 
to 
$Q(\veceta \to \veceta')$ in (\ref{heat}), where 
\begin{eqnarray} 
 \jt(\veceta \to \veceta')
&\equiv&\sum_i[\eta_i(1-\eta'_{i})
\eta'_{i+(0,1)}(1-\eta_{i+(0,1)})\nonumber \\
&-&\eta'_i(1-\eta_i)\eta_{i+(0,1)}
(1-\eta'_{i+(0,1)})].
\label{jt}
\end{eqnarray} 
Note that in the $x$-direction, the particles are still driven by $E$.    
Then, the averaged current $\bar J_\epsilon$  in the $y$-direction 
is defined as   
\begin{equation}
\bar J_\epsilon\equiv \frac{1}{L}\bra \jt(\veceta \to \veceta')\kets^{E,\epsilon}.
\end{equation}
Using this  $\bar J_\epsilon$, the conductivity $\sigma$ is written 
\begin{equation}
\sigma\equiv \lim_{\epsilon\to 0}  \frac{\bar J_\epsilon}{\epsilon}. 
\label{sigdef}
\end{equation}
In the inset of Fig. \ref{fig:ein},  $\sigma$ is plotted as a function 
of the system  size, $L$, in the cases $E=0$ and $E=10$ with  
$\rho=0.5$.  Note that the values of $\sigma$ in the 
case $E=0$ are smaller than  those in the case $E=10$, and that 
qualitatively, this difference is not the same as that seen for  $D$.

We previously measured the intensity of density 
fluctuations $\chi\equiv L\ell(\bra\rho_{\ell}^2\kets^E-
(\bra\rho_{\ell}\kets^E)^2)$, where $\rho_{\ell}\equiv 
\sum_{i\in \Omega_{\ell}}\eta_i/|\Omega_{\ell}|$ and 
$\Omega_\ell=\{(i_x,i_y)|1\le i_x\le L, 
L/2-\ell/2-1 \le i_y\le L/2+\ell/2\}$. (See Fig. 4 in  Ref. \cite{HSI}.)
Note that $\ell$ is chosen so that it satisfies $\xi \ll \ell \ll L$ 
where $\xi$ is a correlation length.  Using these values of $\chi$,  
in Fig. \ref{fig:ein}, we plot $\sigma T$ as 
a function of $D\chi$ in the cases  $(\rho,E)=(0.5,10),(0.4,10),
(0.3,10),(0.5,3)$ and $(0.5,0)$. Noting that the thin-dotted line 
represents $D\chi=\sigma T$,  we find that even though the values 
of $D$, $\sigma$ and $\chi$  are different  from those in the 
equilibrium case,  along the direction transverse  to the driving force 
$E$,  the Einstein relation (\ref{ein}) is valid, within the precision of 
the numerical computations.

\paragraph*{Fluctuation-response relation:} 

%
\begin{figure}
\begin{center}
\includegraphics[width=8cm]{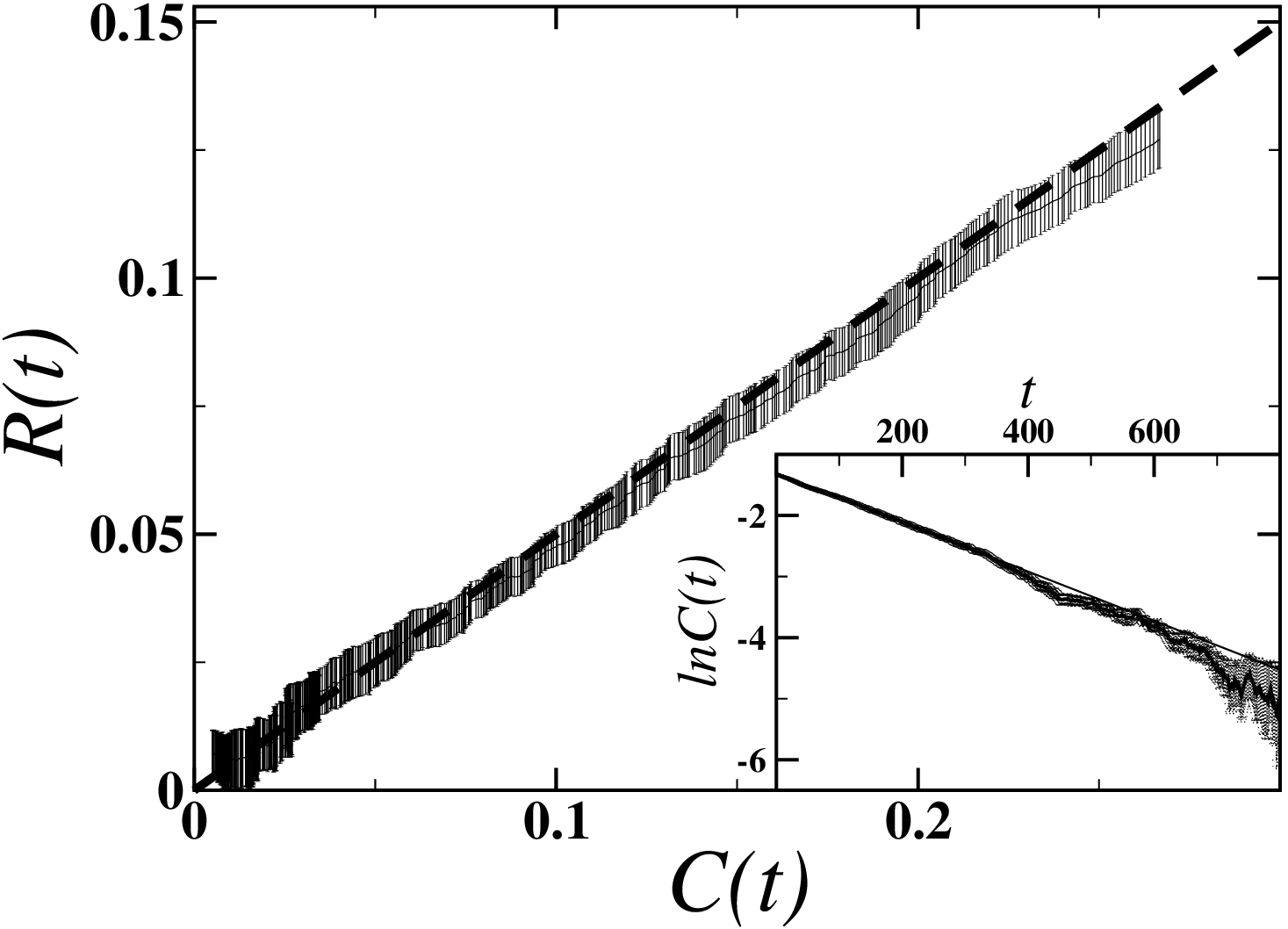}
\end{center}
\caption{$R(t)$ plotted as a function of $C(t)$ in the case 
$(\rho,E)=(0.5,10)$, with $L=32$ and $\Delta=-0.2$ over the interval  
 $0\le t\le800$ MCS.  The  line represents $R(t)=\beta C(t)$,  
where $T=2 (\beta=0.5)$. The error bars represent these of $R(t)$,  
because the uncertainty on $C(t)$ is smaller. In the inset, 
$\ln C(t)$ is plotted. The line there represents $-0.00403t-1.31$. 
}
\label{fig:rc}
\end{figure}

We next study in the fluctuation-response 
relation, which is also a representative universal relation in the 
linear response theory. Again in this case, we focus on 
the properties of the system 
in the direction transverse to the driving force $E$. 

First, we employ the same procedure as in  the measurement of $D$ 
to introduce a time dependent response function $R(t)$.  That is, 
we prepare the steady state under  the perturbation $V(i_y)$  and then 
remove this perturbation at $t=0$. 
Because the profile of the coarse grained density $\rho(i_y)$ 
is changed by the removal of $V(i_y)$, we  make this change explicit by 
defining $R(t)$ in the following form:  
\begin{equation}
R(t)\equiv-\frac{\bra\hat\rho(t)\ket^{V}_{E}}{L\Delta}. 
\label{res}
\end{equation}
We remark that the decaying behavior of $R(t)$ in the case $E=10$ 
is plotted as that of $-\bra\hat\rho(t)\ket_{E}^{V}/
\Delta$ in Fig. \ref{fig:exp}.  

Next, we introduce the time correlation function of density fluctuations   
in the direction transverse to the driving force: 
\begin{equation}
C(t)\equiv \bra\hat\rho(t)\hat\rho(0)\kets^{E}. 
\end{equation}
In the inset of Fig. \ref{fig:rc}, $\ln C(t)$ is plotted as a function 
of time in the case $E=10$, with $\rho=0.5$ and $L=32$. 
For the NESS, $C(t)$, like $R(t)$  
decays exponentially in time.  

Here, in the equilibrium case ($E=0$), using $R(t)$ and $C(t)$, 
the fluctuation-response relation is given by 
\begin{equation}
C(t)=T R(t). 
\label{frr}
\end{equation} 
In the NESS far from equilibrium studied here, 
because 
$R(t)$ and $C(t)$ exhibit 
 a similar behavior, and because the fluctuation relation,  
which is essentially the same as $C(0)=T R(0)$,  
has previously been found to hold \cite{HSI}, we conjecture  
that  (\ref{frr}) is valid.  
 
To demonstrate its  validity explicitly, in Fig. \ref{fig:rc}, in the case 
$E=10$ with $\rho=0.5$ and $L=32$, $R(t)$ is plotted as a function 
of  $C(t)$ over the interval $0\le t\le 800$ MCS. It is seen that 
the slope is equal to $1/T$, within the precision of the 
numerical computations.

\paragraph*{Green-Kubo relation:}  

%
\begin{figure}
\begin{center}
\includegraphics[width=8cm]{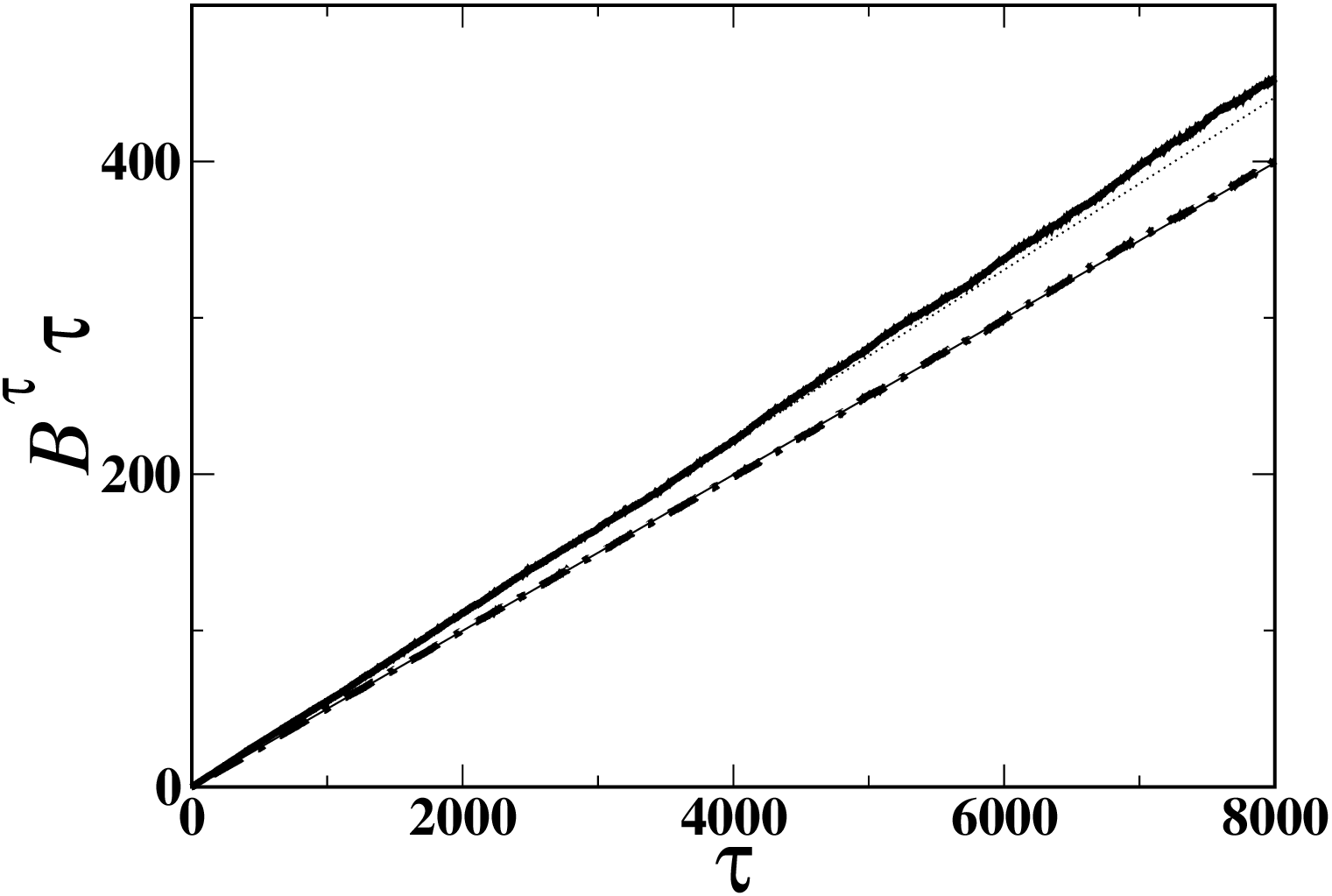}
\end{center}
\caption{$B^\tau \tau$  plotted as a function of $\tau$ in the case 
$(\rho,E)=(0.5,0)$ (the dotted line) and $(\rho,E)=(0.5,10)$ (the 
straight line) with $L=32$.  The thin  line represents 
$0.050\tau$ and the thin dotted line represents $0.055\tau$. In the 
case $E=10$, it is seen that the line deviates from the thin 
dotted line at large $t$ MCS.  
}
\label{fig:btau}
\end{figure}

Finally, again considering 
 the properties along the direction 
transverse to the external 
driving force (the $y$ direction),  we investigate the validity 
of the Green-Kubo relation for NESS far from equilibrium. 
 
Using the spatially-averaged current in the direction transverse 
to the  external driving force, $\jt$, defined in (\ref{jt}),  
we begin by  the $\tau$ dependent current $J^\tau$, which represents 
the net number of particles that move in the $y$ direction during a time 
of $\tau$ MCS,    
\begin{equation}
J^\tau\equiv \frac{1}{\tau L^2}\sum_{k=1}^{\tau L^2}
\jt(\veceta(k-1)\to\veceta(k)). 
\end{equation}
Then, using this expression for  $J^\tau$, 
the intensity of the current fluctuations is defined by
\begin{equation}
B^{\tau}\equiv\frac{\tau L^2}{2}\bra (J^\tau)^2\kets^{E}. 
\end{equation}
In Fig \ref{fig:btau}, $B^\tau \tau$ is plotted as a function of 
$\tau$ in the cases $E=0$ and $E=10$, with $\rho=0.5$ and $L=32$,  
respectively.   
It is seen that in the case $E=10$, the line fitted for small times  
(but much larger than the relaxation time of the current 
correlations) deviates slightly for large times, while in the case $E=0$, 
$B^\tau \tau$ and $0.050\tau$ are equal within the numerical precision 
for all times.  This bending 
behavior of $B^\tau\tau$ might reflect the effect of 
a long time tail in this NESS. 

\begin{figure}
\begin{center}
\includegraphics[width=6cm]{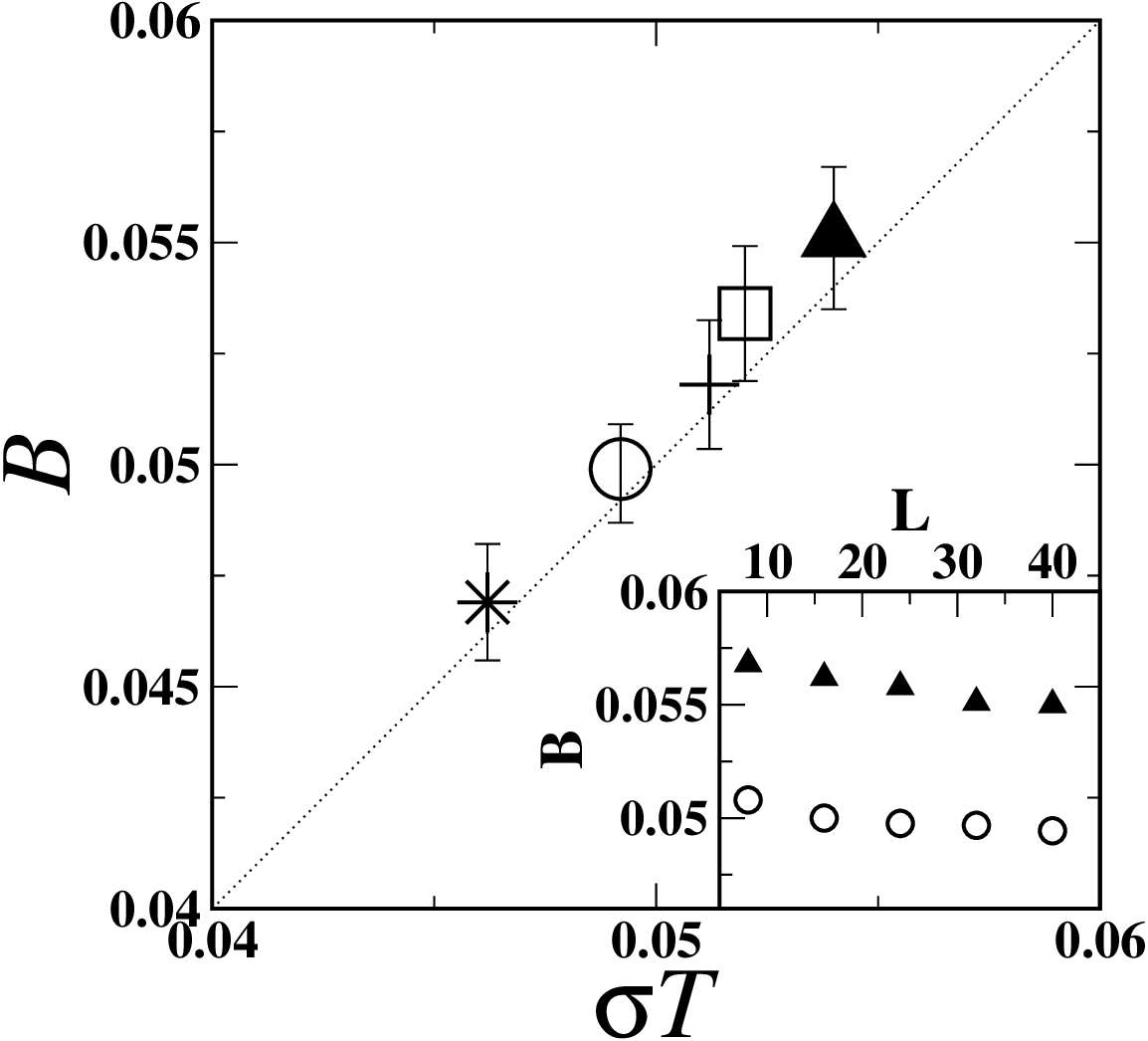}
\end{center}
\caption{$\sigma T$ as a function of $B$. 
The triangle, square, star, plus, and circle correspond to 
$(\rho,E)=(0.5,10),(0.4,10),(0.3,10),(0.5,3)$ and $(0.5,0)$. 
The inset displays the $L$ dependence of $B$. The triangles 
and circles correspond to $(\rho,E)=(0.5,10)$ and  $(0.5,0)$, 
respectively.  
}
\label{fig:gk}
\end{figure}

In the case $E=0$, defining $B$ as the slope of $B^\tau \tau$, the 
Green-Kubo relation \cite{kubo}  can be written 
\begin{equation}
B=\sigma T. 
\label{gk}
\end{equation}
In the case $E\ne 0$,  we define $B$ as  the slope of $B^{\tau}\tau$ 
obtained from the fitting in the early time regime. With this 
definition, the size dependence of $B$ in the cases $E=0$ and $E=10$ 
with $\rho=0.5$ is plotted in the inset of Fig. \ref{fig:gk}.  
The difference between the values of $B$ in the cases   
$E=0$ and $E=10$ is qualitatively the same as that for $\sigma$.   

Considering this similarity between $\sigma$ and $B$, in Fig. 
\ref{fig:gk},  we plot $\sigma T$ as a function of $B$  in the 
cases $(\rho,E)=(0.5,0), (0.5,10),(0.5,3),(0.4,10),(0.3,10)$ with 
$L=32$. The Green-Kubo relation (\ref{gk}) 
is valid for the NESS considered here. 
However, the deviation seen in Fig. \ref{fig:gk} is  somewhat 
larger than that in Fig. \ref{fig:ein}.  

%
%

\paragraph*{Summary:}  

In this paper, we have reported the results of numerical experiments on 
the two-dimensional DLG focusing on the properties along the direction 
transverse to the external driving force $E$.  We find that the Einstein 
relation, the fluctuation-response relation and the Green-Kubo relation 
hold in the NESS far from equilibrium.  (Note that the validity of the  
fluctuation relation for such a state was previously demonstrated  
in Ref. \cite{HSI}.)  

Compared with this validity of relations in the direction transverse 
to the external driving force, we remark the properties along the  
direction parallel to the external driving force.  We studied a 
one-dimensional DLG, and found that the phenomena observed along the  
direction parallel  to the external driving force 
seemed to be more complicated than those observed along the 
direction transverse to  the external driving force \cite{HSIV}. 

We end with some discussion of the detailed balance of fluctuations, 
which has a deep connection with  the validity of universal  relations 
in the linear response regime near equilibrium states.  In our DLG, 
the detailed balance  condition for $c(i,j;\veceta)$ does not hold in 
the case $E\ne 0$. However, the numerical confirmation of the universal 
relations presented here suggests  the detailed balance of 
macroscopic fluctuations.  We point out  that, with regard to this 
topic,   D. Gabielli et al. studied a stochastic model for which  the 
detailed balance condition does not hold, and derived the Onsager's 
reciprocity, which is also the linear response relations for macroscopic 
quantities \cite{gabli}.

The author acknowledges S. Sasa and  H. Tasaki for discussions on NESS. 
This work was supported by a JSPS Research Fellowship  for Young 
Scientists (Grant No. 1711222).


\begin{thebibliography}{10}
\bibitem{chemi}
This chemical  potential was defined  operationally.  
The theoretical background employed in  \cite{HSI} is  
supported  by the result of Ref. \cite{sst}.
\end{thebibliography}
\end{document}